\documentclass[aps,prd,nopacs,floatfix,notitlepage,superscriptaddress,nofootinbib,twocolumn,a4paper]{revtex4-1}
\usepackage{amsfonts,amsmath,units,wasysym,epsfig,graphicx,verbatim,color,subfigure,graphicx,bm,mathrsfs,lipsum,hyperref}
\usepackage[utf8]{inputenc}
\usepackage[normalem]{ulem}  

\begin{document}

\newcommand{\USAL}{Departamento de F\'isica Fundamental, Universidad de Salamanca, Plaza de la Merced S/N, E-37008 Salamanca, Spain}

\newcommand{\Uliege}{Space Sciences, Technologies and Astrophysics Research (STAR) Institute, Universit\'e de Li\`ege, B\^at. B5a, 4000 Li\`ege, Belgium}

\newcommand{\brussels}{Institut d’Astronomie et d’Astrophysique, Universit\'e Libre de Bruxelles, CP 226, B-1050 Brussels, Belgium}

\title{Exploring the limits of nucleonic metamodelling using different relativistic density functionals}

\author{Prasanta Char}
\email{prasanta.char@usal.es}
\affiliation{\USAL}\affiliation{\Uliege}

\author{Chiranjib Mondal} 
\email{chiranjib.mondal@ulb.be} 
\affiliation{\brussels}

\begin{abstract}
In this work, we explore two classes of density dependent relativistic mean-field models, their predictions of proton fractions at high densities and neutron star structure. We have used a metamodelling approach to these relativistic density functionals. We have generated a large ensemble of models with these classes and then applied constraints from theoretical and experimental nuclear physics and astrophysical observations. We find that both models produce similar equation of state and neutron star mass-radius sequences. But, their underlying compositions, denoted by the proton fraction in this case, are vastly different. This reinstates previous findings that information on composition gets masqueraded in $\beta$-equilibrium. Additional observations of non-equilibrium phenomena are necessary to pin it down.

\end{abstract}
\maketitle

\section{Introduction} 
Dense cores of neutron stars (NSs) present a unique laboratory to study the behavior of matter at densities that are a few times the normal nuclear matter densities. The composition and properties of such matter which is described mathematically by its equation of state (EOS) is still not known precisely. The first major astrophysical constraints on the NS matter came with the radio observations of massive pulsars in the previous decade and subsequent refinement of the measurements \cite{Demorest:2010bx,Antoniadis:2013pzd,Fonseca:2016tux, Arzoumanian:2017puf,Cromartie:2019kug,Fonseca:2021wxt}. The multimessenger gravitational wave event, GW170817 and its electromagnetic counterparts have also brought forth plethora of information regarding nuclear processes, material abundances in the universe \cite{TheLIGOScientific:2017qsa,LIGOScientific:2017ync}. It has provided the testbeds for fundamental physical theories like general relativity, nuclear physics. At the same time, NICER collaboration has reported several simultaneous mass-radius measurements from X-ray-emitting NSs that can potentially constrain the NS EOS \cite{Riley:2019yda,Miller:2019cac,Riley:2021pdl,Miller:2021qha,Choudhury:2024xbk}. 

Direct calculation of the NS EOS from quantum chromodynamics (QCD) is not feasible because the QCD couplings exhibits a peculiar behavior of being strong at nuclear densities, but weak at very high densities and temperature. Hence, instead of the proper QCD description of dense nuclear matter, effective field theory approaches are very popular and widely used to understand the properties of both infinite nuclear matter and finite nuclei \cite{Glendenning:1997wn}. In most of these models, effective interactions are usually employed in the mean-field approximation. The parameters of the interactions are not computed from any fundamental theory, rather fitted to 
experimental and observational properties. Relativistic mean field (RMF) models are the class of phenomenological models that provides an energy density functional to calculate the EOS of dense matter \cite{Walecka:1974qa,Serot:1984ey}. RMF models have been successful in describing the properties of nuclear matter at nuclear saturation density ($n_{sat}$) and many properties of finite nuclei \cite{Boguta:1977xi}. They can also be extended for situations like high densities and large charge asymmetry that may arise in the neutron star mergers and gravitational core-collapse \cite{Oertel:2016bki}. These are traditionally the meson exchange models with Dirac spinors where the mesons are effective fields to capture the essence of the strong interaction \cite{Dutra:2014qga,Sun:2023xkg}. 

A newer class of RMF models was proposed to include density dependent meson-baryon couplings instead of adding extra terms in the Lagrangian to explain the high density behavior \cite{Fuchs:1995as,Lenske:1995wyj,Hofmann:2000vz,Hofmann:2000mc}. The self energy calculated within Dirac-Brueckner-Hartree-Fock approximation for realistic nucleon-nucleon interaction is used to optimize the functional form of the density-dependent couplings. Several forms of density dependence have been proposed. In this work, we will study two of them to understand their scope in describing different astrophysical situations. The first of them, proposed by \textcite{Typel:1999yq} will be denoted by TW, and the second one proposed by \textcite{Gogelein:2007qa} will be denoted by GDFM, hereafter in this article. The TW functional has been successful in describing both ground state and collective excitation properties of several hundreds of finite nuclei measurements over the whole nuclear chart \cite{Pena-Arteaga:2016clz, Lalazissis:2019oey, Taninah:2024eqe}. It has also been applied to astrophysical simulations of core-collapse supernova and binary neutron star mergers \cite{Hempel:2011mk,Banik:2014qja}. Different variations of TW model have been realised to explore the finite nuclei properties \cite{Typel:2018cap}. GDFM functional has also been used to understand NS crust properties \cite{Gogelein:2007qa,Avancini:2008kg}. Recently, it was explored as a viable candidate for a relativistic metamodel for application to NS properties \cite{Char:2023fue,Scurto:2024ekq}. Metamodelling approaches are useful to understand and incorporate systematically the uncertainties of model parameters. In the context of dense matter, a nucleonic metamodel has been developed recently, to incorporate the uncertainties of nuclear matter parameters (NMPs), to study the parameter space, and to interpolate between existing parameter sets \cite{Margueron:2017eqc,Margueron:2017lup,Margueron:2018eob}. We will apply this metamodelling technique to relativistic density functionals to understand their strength in generating sets of NMPs within the current ranges of experimental uncertainties. We will also study both the high- and low-density extrapolations of those functionals to achieve diverse realizations of compositions inside neutron stars (NS). The EOS at subsaturation density and a little above saturation is constrained from the experimental estimations of the NMPs and the theoretical calculations of symmetric nuclear matter(SNM) and pure neutron matter (PNM) properties from chiral effective field theory ($\chi$-EFT) \cite{Tews:2012fj,Hebeler:2013nza,Lynn:2015jua, Drischler:2017wtt, Carbone:2019pkr,Leonhardt:2019fua,Huth:2020ozf}. Some perturbative QCD calculations also exist at very high density ($\sim 40 n_{sat}$) and may have some consequence for the NS matter \cite{Gorda:2022jvk}. Relativistic density functionals have been investigated to find the most optimized parametrizations incorporating a large set of different experimental and observational information \cite{Traversi:2020aaa,Malik:2022zol,Zhu:2022ibs,Beznogov:2022rri,Malik:2023mnx,Salinas:2023nci,Huang:2023grj,Providencia:2023rxc,Li:2024pts}. 

Recently, machine learning techniques have been employed to infer the composition of neutron star matter from observations, using RMF equations of state for training \cite{Carvalho:2023ele,Carvalho:2024kgf}. If the compositions of the EOSs for those training datasets are constrained by the construction of the RMF functionals, their predictions could be biased. In particular, most RMF functionals are known to provide a narrow range of proton fraction \cite{Malik:2022zol,Malik:2023mnx}. In contrast, the GDFM functional is shown to have higher proton fractions \cite{Char:2023fue,Scurto:2024ekq}. Therefore, a comparison among the capabilities of RMF models is essential to create robust datasets for training. This is one the main motivation of the present work where we test the standard TW functional against the GDFM functional, within a metamodelling approach. We compare them systematically under identical conditions to evaluate their capabilities.

The structure of the paper is the following. First, we provide an introduction to density-dependent relativistic hadronic models. After that, we explain metamodelling of relativistic functionals in Bayesian analysis with the details of the priors and the constraints for the posteriors. Finally, we discuss our results and outline our conclusions.

\section{Formalism}
\subsection{Relativistic model}
The Lagrangian density used in the present work is of the form
\begin{widetext}
\begin{eqnarray}
\mathcal{L}_{\rm DD} &=& \overline{\psi}(i\gamma^\mu\partial_\mu - M)\psi 
+ \Gamma_\sigma(n_B)\sigma\overline{\psi}\psi 
- \Gamma_\omega(n_B)\overline{\psi}\gamma^\mu\omega_\mu\psi 
-\frac{\Gamma_\rho(n_B)}{2}\overline{\psi}\gamma^\mu{\boldsymbol \rho}_\mu \cdot {\boldsymbol \tau}
\psi  \nonumber \\
&+& \frac{1}{2}(\partial^\mu \sigma \partial_\mu \sigma - m^2_\sigma\sigma^2)
- \frac{1}{4}F^{\mu\nu}F_{\mu\nu} + \frac{1}{2}m^2_\omega\omega_\mu\omega^\mu 
-\frac{1}{4}\vec{B}^{\mu\nu}\vec{B}_{\mu\nu}+\frac{1}{2}m^2_\rho
{\boldsymbol \rho}_\mu \cdot {\boldsymbol \rho}^\mu .
\label{dldd}
\end{eqnarray}
\end{widetext}
Here, $\sigma,\,\omega_{\mu}$,and $\boldsymbol{\rho_{\mu}}$ are isoscalar-scalar, isoscalar-vector and isovector-scalar effective meson fields, respectively, mediating the strong interaction among the nucleons, represented by the field $\psi$.
{The strength of the interactions between the mesons and the nucleons is determined by the coupling constants $\Gamma$, which are density dependent. We have considered two different types of density dependence for comparison, namely, TW \cite{Typel:1999yq} and GDFM \cite{Gogelein:2007qa}. The density dependence for the TW type is given by,}
\begin{eqnarray}
\Gamma_i(n_B) &=& \Gamma_i(n_{sat})f_i(x),\quad\mbox{with}\quad \\
f_i(x) &=& a_i\frac{1+b_i(x+d_i)^2}{1+c_i(x+d_i)^2},
\label{gamadefault}
\end{eqnarray}
for $i=\sigma,\omega$, and
\begin{eqnarray}
\Gamma_\rho(n_B)=\Gamma_\rho(n_{sat})e^{-a(x-1)},\quad\mbox{with}\quad x=n_B/n_{sat}.
\end{eqnarray}
Furthermore, these parameters are not independent. Additional conditions are used for them as: $f_i(1)=1, f_{i}''(0)=0$ and $f_{\sigma}''(1) = f_{\omega}''(1)$. The GDFM type density dependence is described as \cite{Gogelein:2007qa}
\begin{equation}
 \Gamma_i(n_B)=a_i+(b_i+d_i\,x^3)e^{-c_i\,x}, \quad\mbox{with}\quad x=n_B/n_0,
\label{GDFM}
\end{equation}
where $n_0$ is a constant scaling density, different from $n_{sat}$. 
The GDFM functional has been used to study the properties of neutron stars recently in Refs. \cite{Char:2023fue,Scurto:2024ekq}.

\begin{table}
    \centering
    \begin{tabular}{c|c}
         \hline
         \hline 
         Parameter& Assumed range of values  \\
         \hline
         \hline
         $ n_{sat}$ (fm$^{-3}$)  & 
         $0.14,0.17$\\
         $ E_{sat}$ (MeV) & 
         $-17,-14$ \\
         $ K_{sat}$ (MeV) & 
         $150,350$ \\
         $ E_{sym}$ (MeV) & 
         $20,45$ \\
         $ L_{sym}$ (MeV) & 
         $20,180$ \\
         \hline
         \hline
    \end{tabular}
    \caption{Ranges of values considered for various nuclear empirical parameters {to construct the priors of relativistic density functional models.}}
    \label{tab:prior}
\end{table}
\subsection{Bayesian analysis}\label{sec:bayes_methodology}
We perform a Bayesian analysis for TW and GDFM models by applying nuclear and astrophysical constraints. {The details of the techniques used can be found in Refs. \cite{universe7100373,Char:2023fue,Mondal:2022cva}. For completeness, we repeat a few key aspects and also highlight additional changes made.} Simultaneous generation of unified EOSs and solving the TOV equations to apply the astrophysical constraints require large computational resources.
This is why we broke down the analysis in two steps. First, we construct the likelihood for the NMPs at saturation calculated from the model parameters. We accept any parameter set in our calculation if the lower order NMPs fall within the range described in Tab. \ref{tab:prior}. We also calculated the SNM and PNM properties in the non-relativistic formalism \cite{Margueron:2017eqc} to apply the $\chi$-EFT constraints. Unlike our previous calculation \cite{Char:2023fue}, where we used a pass-band type filter for the $\chi$-EFT filter at discrete  density points, we have implemented the same here as independent Gaussian distributions at those density points.
In the same setting as Ref. \cite{Char:2023fue}, this comes down to assigning the probabilities as a modified Gaussian distribution \cite{Scurto:2024ekq},

   \begin{eqnarray} \label{eq:gaussian}
        &&P(\chi\text{EFT}|{\mathbf X})\\ \nonumber
        && \;\;\;\propto \prod_{i =1}^{N} \begin{cases} P_U^i(x_i) & \mbox{if } x_{min}^i < x_i({\bf X}) < x_{max}^i \\ P_G^i(x_i) & \mbox{otherwise} \end{cases} ,
    \end{eqnarray}
where, $\mathbf{X}$ represents the RMF model parameters and N is the number of discrete points where the filter was applied and $x$ in our case were the energy per particle of SNM and PNM. In this way, we kept all the parameter sets those satisfy constraints of Tab. \ref{tab:prior} with associated probabilities. 
The quantities $x_{min}^i$ and $x_{max}^i$ are the lower and upper bound of the theoretical band at the same density point, and where
    \begin{equation}
        P_U^i(x_i)=\frac{0.682}{2\sigma_i},
    \end{equation}
    while the Gaussian function is defined as
    \begin{equation}
        P_G^i(x_i)=\frac{1}{\sigma_i\sqrt{2\pi}}e^{-\frac{1}{2}(\frac{x_i-\mu_i}{\sigma_i})^2},
    \end{equation}
    with the mean, $\mu_i=(x^i_{max}+x^i_{min})/2$ and the standard deviation, $\sigma_i=(x^i_{max}-x^i_{min})/2$.
In this way, we do not immediately discard all the models that are outside the theoretical limit.
Then we have used a Nested Sampling method implemented in {\tt PyMultiNest} software \cite{Buchner:2014nha} to find the equally weighted posteriors for model parameters. We have also ensured the optimal sampling of the parameter space in this process. {We define our priors for these two models by further optimization of these NMPs with the AME2016 nuclear mass table \cite{Wang:2017fhd}. Hence, these priors carry combined constraints coming from theoretical and experimental nuclear physics, thus can be considered as nuclear physics-informed priors.}
\begin{table}
    \centering
    \begin{tabular}{ccc}
    \hline
    \hline
          Parameters & Maximum value & Minimum value \\
         \hline 
        \hline
        & GDFM &   \\ 
        \hline 
        \hline       
        $a_\sigma$ & 10.295748 & 6.9837231 \\
        $b_\sigma$ & 3.2618188 & 2.0238622 \\
        $c_\sigma$ & 2.7911622 & 1.6943625 \\
        $d_\sigma$ & 5.2779045 & 2.4805772 \\
        $a_\omega$ & 13.6596588 & 9.1064392 \\
        $b_\omega$ & 2.35939872 & 1.57293248 \\
        $c_\omega$ & 8.2559356 & 5.0097963 \\
        $d_\omega$ & 1.6719065 &  0.67148104 \\
        $a_\rho$ & 1.0 & -1.0 \\
        $b_\rho$ & 7.312709592 & 4.875139728 \\
        $c_\rho$ & 0.66405387 & 0.40285884 \\
        $d_\rho$ & 1.2092027 & -1.2112768 \\
        \hline 
        \hline
        & TW & \\
        \hline 
        \hline
        $\Gamma_\sigma$ & 10.295748 & 6.9837231 \\
        $\Gamma_\omega$ & 3.2618188 & 2.0238622 \\
        $\Gamma_\rho$ & 2.7911622 & 1.6943625 \\
        $b_\sigma$ & 5.2779045 & 2.4805772 \\
        $c_\sigma$ & 13.6596588 & 9.1064392 \\
        $c_\omega$ & 2.35939872 & 1.57293248 \\
        $a_\rho$ & 8.2559356 & 5.0097963 \\
        \hline
        \hline 
          & subsaturation NMPs & \\
        \hline
        \hline
        $Q_{sat}$ & 1000.0 & -1000.0 \\
        $Z_{sat}$ & 3000.0 & -3000.0 \\
        $Q_{sym}$ & 2000.0 & -2000.0 \\
        $Z_{sym}$ & 5000.0 & -5000.0 \\
        \hline
        \hline 
    \end{tabular}
    \caption{ Ranges of model parameters used to explore the distribution of NMPs {in the Bayesian analysis for GDFM and TW models.}}
     \label{tab:parameters}
\end{table}

After the model parameter space is optimized, we obtain the unified EOS at $\beta$-equilibrium for crust and core. For the crust, we have used the compressible liquid drop approach proposed by Carreau \textit{et. al.} in Ref. \cite{Carreau:2019zdy}. For a given model, the crust can be fully determined by the NMPs in the formalism of Ref. \cite{Carreau:2019zdy}, provided, non-relativistic (NR) metamodel is used. To obtain the crust for the relativistic metamodel fully consistently, one has to solve additional field equations which does not give very different results for the same set of NMPs \cite{Scurto:2024ekq}. This is why we have used the NR metamodel for the crust using the NMPs generated using the RMF parameters, as described in the previous paragraph. At the crust-core junction we match this NR metamodel crust with NR metamodel core, which is matched with the relativistic core at saturation density. This is unlike our previous calculation \cite{Char:2023fue}, where we matched the NR crust at the crust-core junction with the relativistic core, which ensures to have no discontinuities in energy density and pressure for the whole density range. 

\begin{figure*}
    \centering
    \includegraphics[width=0.8\textwidth]{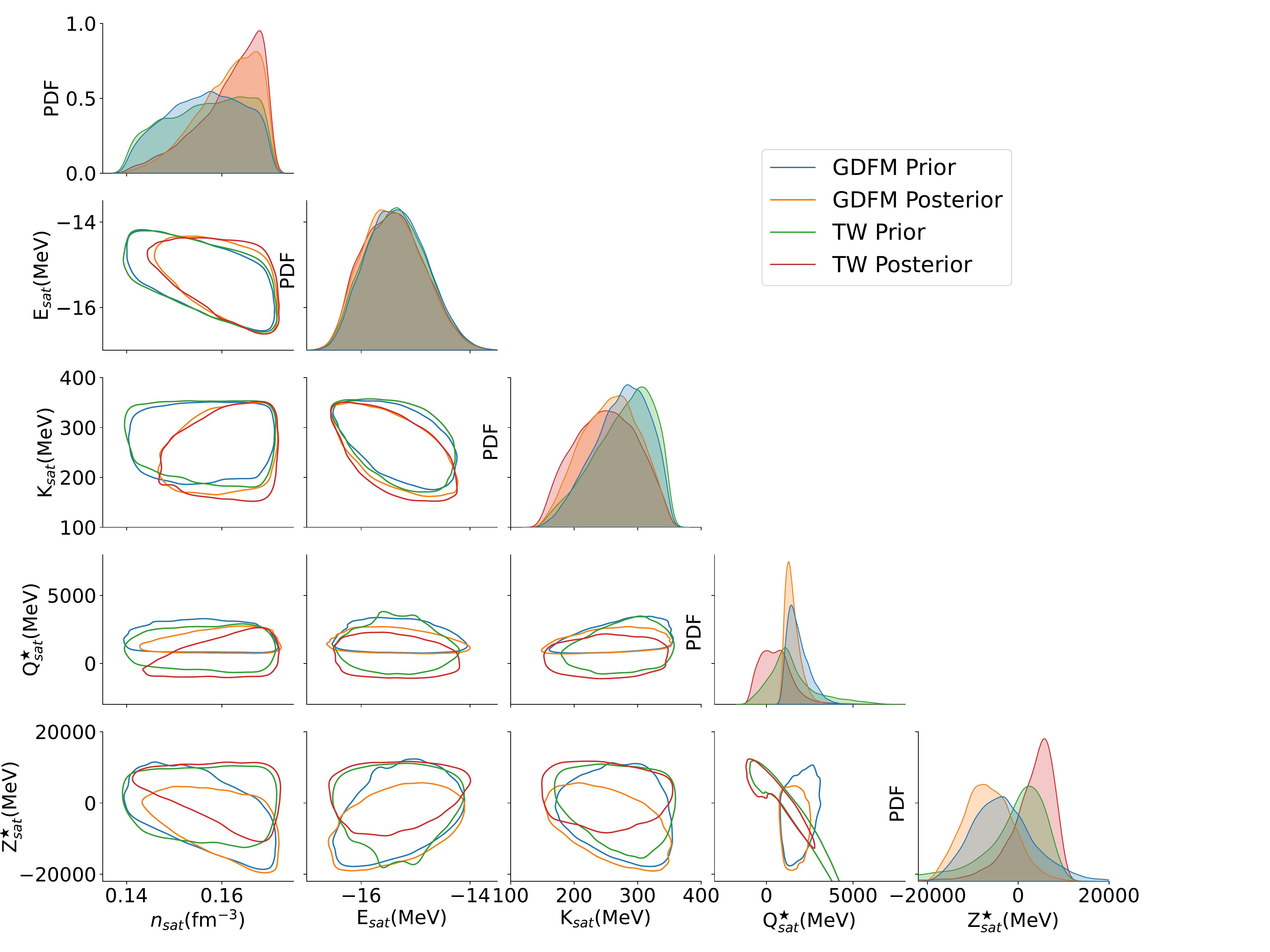}
\caption{Probability distributions of isoscalar NMPs for GDFM and TW, and their correlation contours within the 90\% CI. }
    \label{fig:isoscalar}
\end{figure*}

\begin{figure*}
    \centering
    \includegraphics[width=0.8\textwidth]{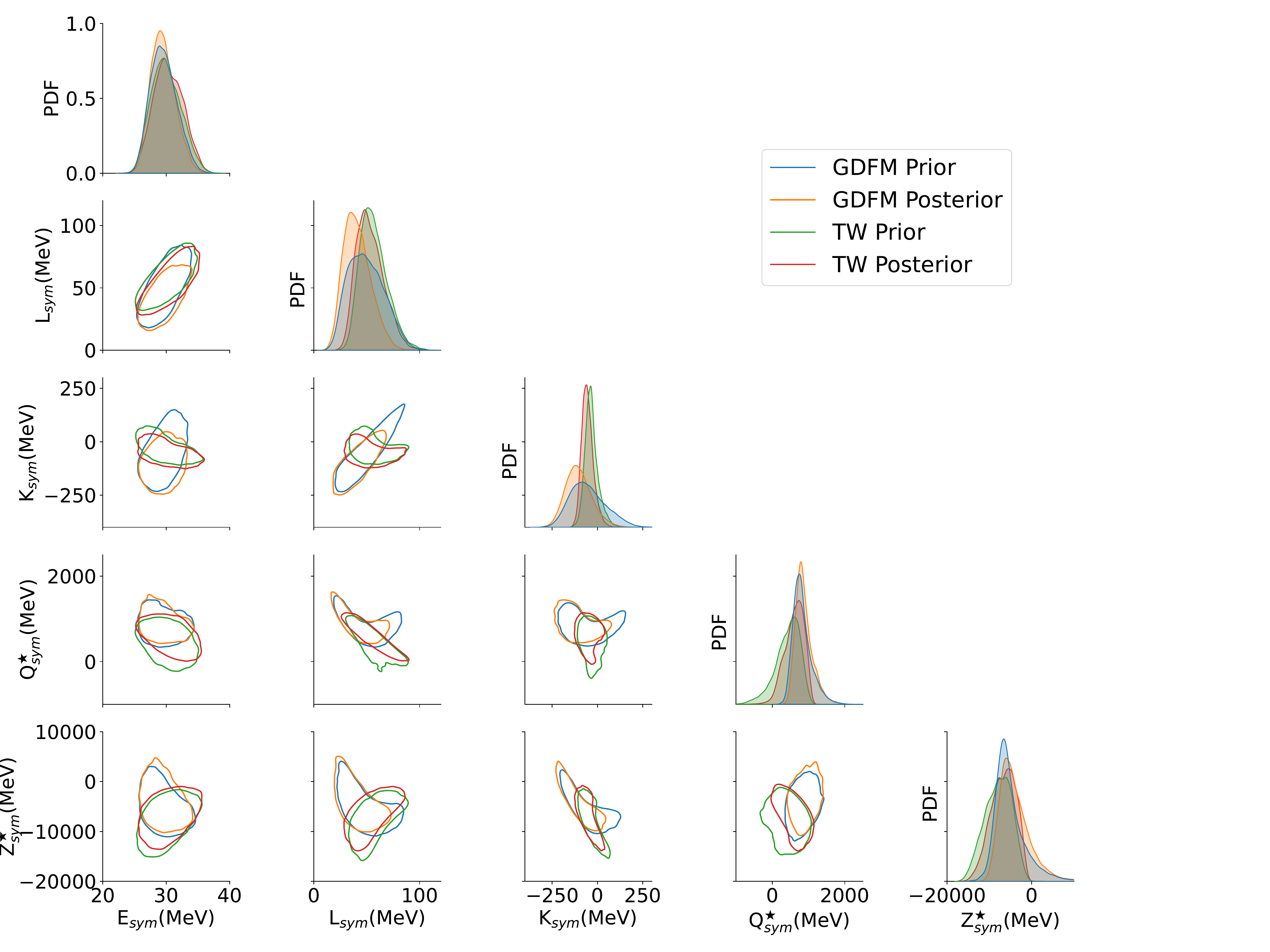}
\caption{The same as Fig. \ref{fig:isoscalar} but for the isovector NMPs. }
    \label{fig:isovector}
\end{figure*}

{We start by generating the NMPs using the GDFM and the TW metamodels by varying the model parameters between the values defined in Table \ref{tab:parameters}. We also provide in the same table the ranges of higher-order parameters $Q_{sat}, Z_{sat}, Q_{sym}$ and $Z_{sym}$ generated independently to calculate the EOS for crust using the NR formalism as explained above. The similar higher-order parameters which are generated from the sampled RMF parameters are referred with an asterisk, \textit{i.e.} $Q_{sat}^*, Z_{sat}^*, Q_{sym}^*$ and $Z_{sym}^*$. They control the high-density behavior of the RMF functionals.}

Once the fully unified EOS is constructed, we apply the astrophysical constraints in the second step of our analysis, following Refs. \cite{universe7100373,Char:2023fue,Mondal:2022cva}. We calculate the mass, radius, and tidal deformability to impose the constraints from pulsar mass observations and tidal deformability estimates from GW170817.  {In this work, we have not imposed the constraints from NICER observations, as we found that almost all of our mass-radius sequences from both metamodels satisfy the $95\%$ CI regions for the NICER sources.} Finally, we use the Bayes theorem to find the posterior distributions of the observables of our interest. {We want to emphasize here that our priors are informed by nuclear physics ($\chi$-EFT and AME masses). The difference between the prior and posterior comes solely from astrophysical constraints.}

\section{Results}

\begin{figure}
    \centering
    \begin{tabular}{c}
     \includegraphics[width=0.5\textwidth]
    {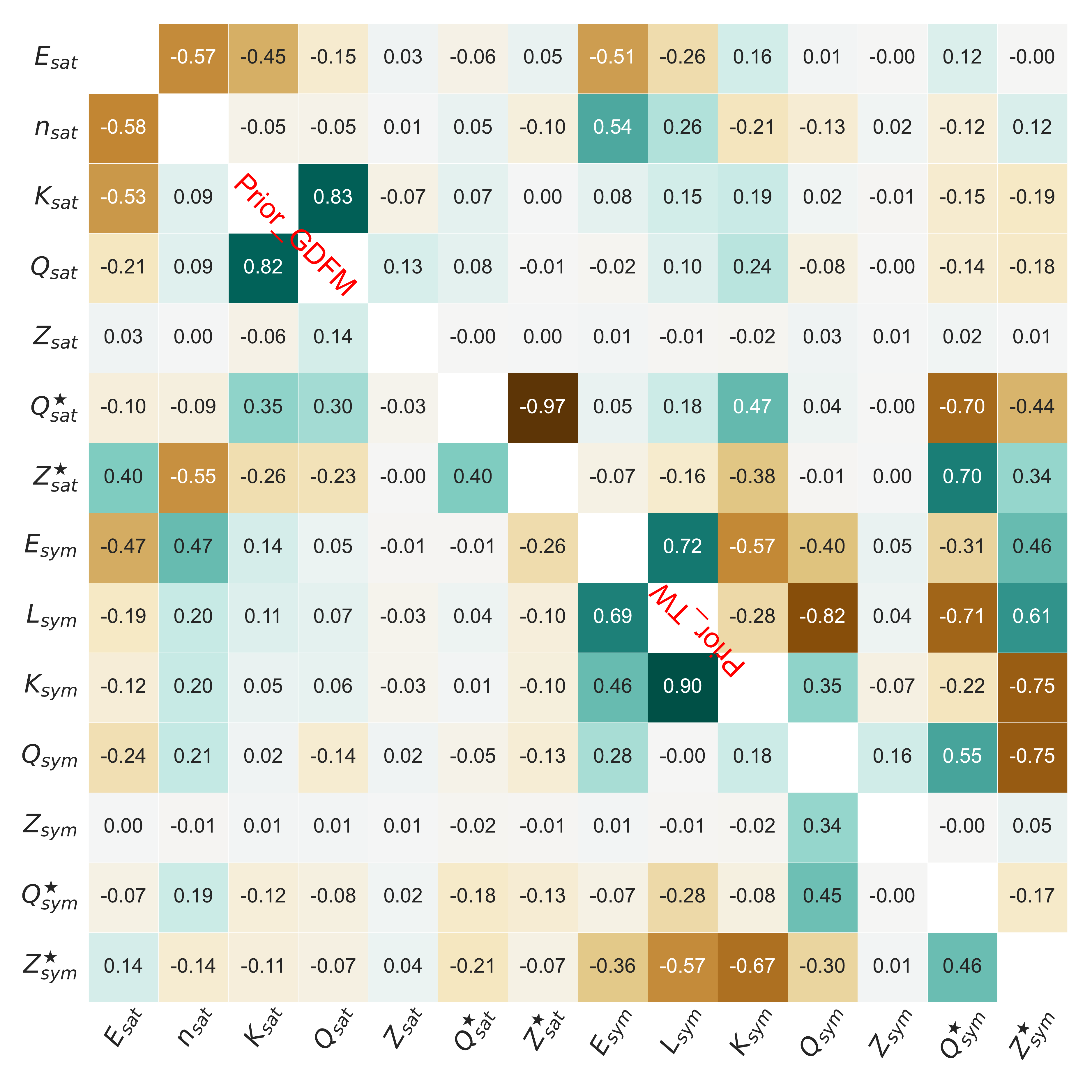} \\
    \includegraphics[width=0.5\textwidth]
    {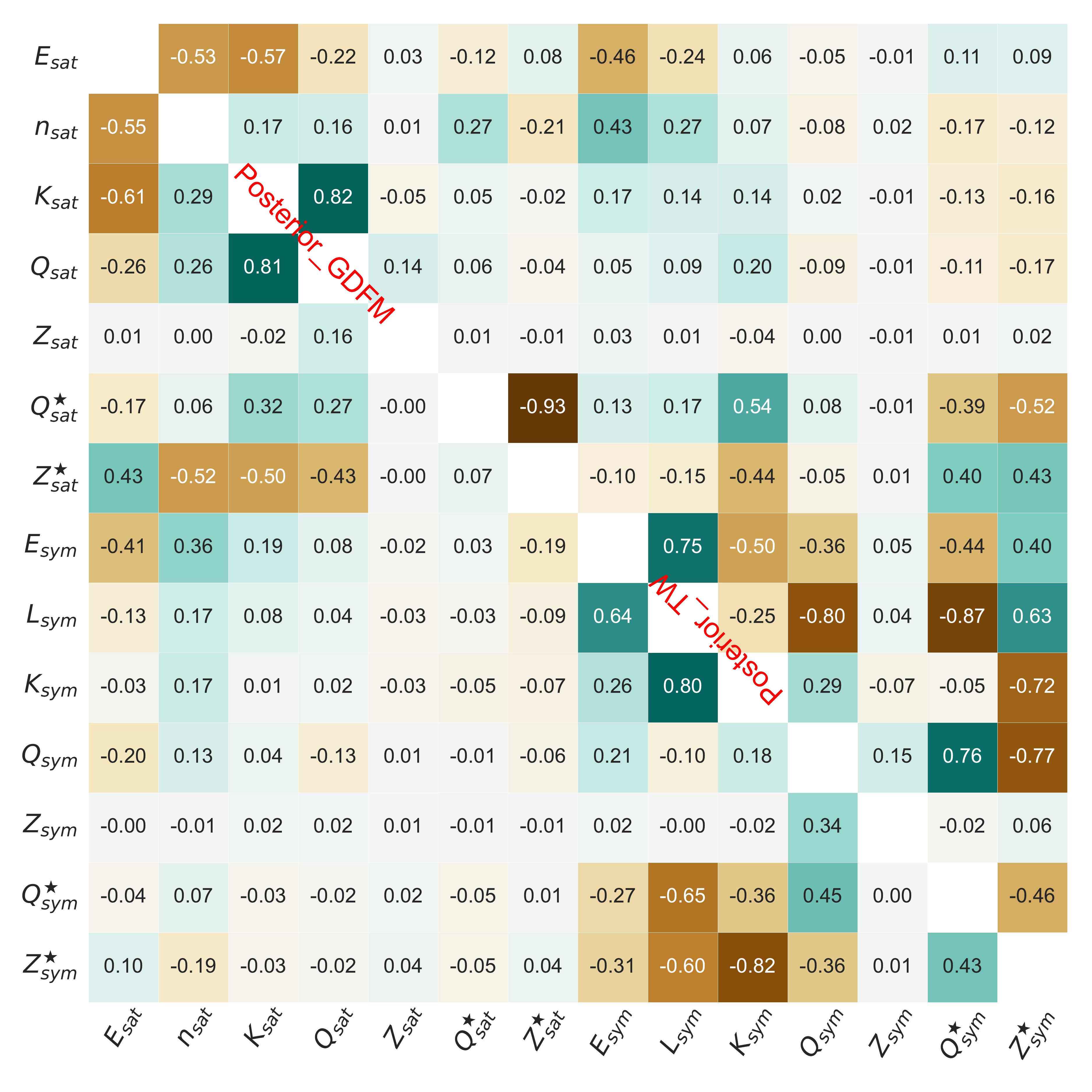}
    \end{tabular}
    \caption{Pearson correlation coefficients between the different NMPs for the GDFM and TW prior (upper panel) and posterior (lower panel) distributions.}
    \label{fig:corr_NMP}
\end{figure}

In Figs. \ref{fig:isoscalar} and \ref{fig:isovector}, we have shown the isoscalar and isovector parameters for the priors and posterios of GDFM and TW, respectively. {In these corner plots, we show the two-dimensional correlated probability distributions and along the diagonal one-dimensional probability distribution functions (PDFs).} From Figure \ref{fig:isoscalar}, we see that the lower order isoscalar parameters of GDFM and TW behave almost identically. Astrophysical constraints do not have any effect on the $E_{sat}$. The $90\%$ CI for $n_{sat}$ reduces marginally from prior to posterior. The incompressibility, $K_{sat}$ also remains unaffected for both GDFM and TW. However, the difference is appreciable for $Q^*_{sat}$ and $Z^*_{sat}$. GDFM does not produce large negative values of $Q^*_{sat}$, but TW extends to larger negative values. On the other hand, GDFM prior has a larger $Z^*_{sat}$ range than TW. The two-dimensional correlation between $Q^*_{sat}$ and $Z^*_{sat}$ shows very different behaviors for GDFM and TW.

In Fig. \ref{fig:isovector}, we see that the distribution of the symmetry energy and its different density derivatives are 
{not drastically different for} GDFM and TW. The astrophysical constraints do not affect {significantly the distribution of lower order symmetry energy parameters.} We found a similar indication in our previous work. The $E_{sym}$ is mostly constrained by the $\chi$-EFT results for the PNM properties. As we consider the higher order isovector parameters, they are affected more by the astrophysical constraints. For example, the GDFM prior for $L_{sym}$ gets reduced. For the TW, the ranges do not change much for the prior and posterior. But, GDFM can produce lower $L_{sym}$ values than TW. When we turn our attention to further higher order isovector NMPs, we find that stark differences emerge from the two models. GDFM produces a larger range for $K_{sym}$. The correlations between $K_{sym}$ and $L_{sym}$ are completely different for GDFM and TW. For $Q^*_{sym}$ and $Z^*_{sym}$, their ranges are comparable, but GDFM produces {larger positive values contrary to TW, which produces more negative values.}  
The reason behind large differences between the higher order parameters can be associated with the parametric freedom that GDFM offers in the $\rho$-meson coupling. Due to this reason, $K_{sym}$, $Q^*_{sym}$, and $Z^*_{sym}$ have different ranges and correlate differently among themselves and {with} $E_{sym}$ and $L_{sym}$.

\begin{figure}
    \centering
    \includegraphics[width=0.5\textwidth]{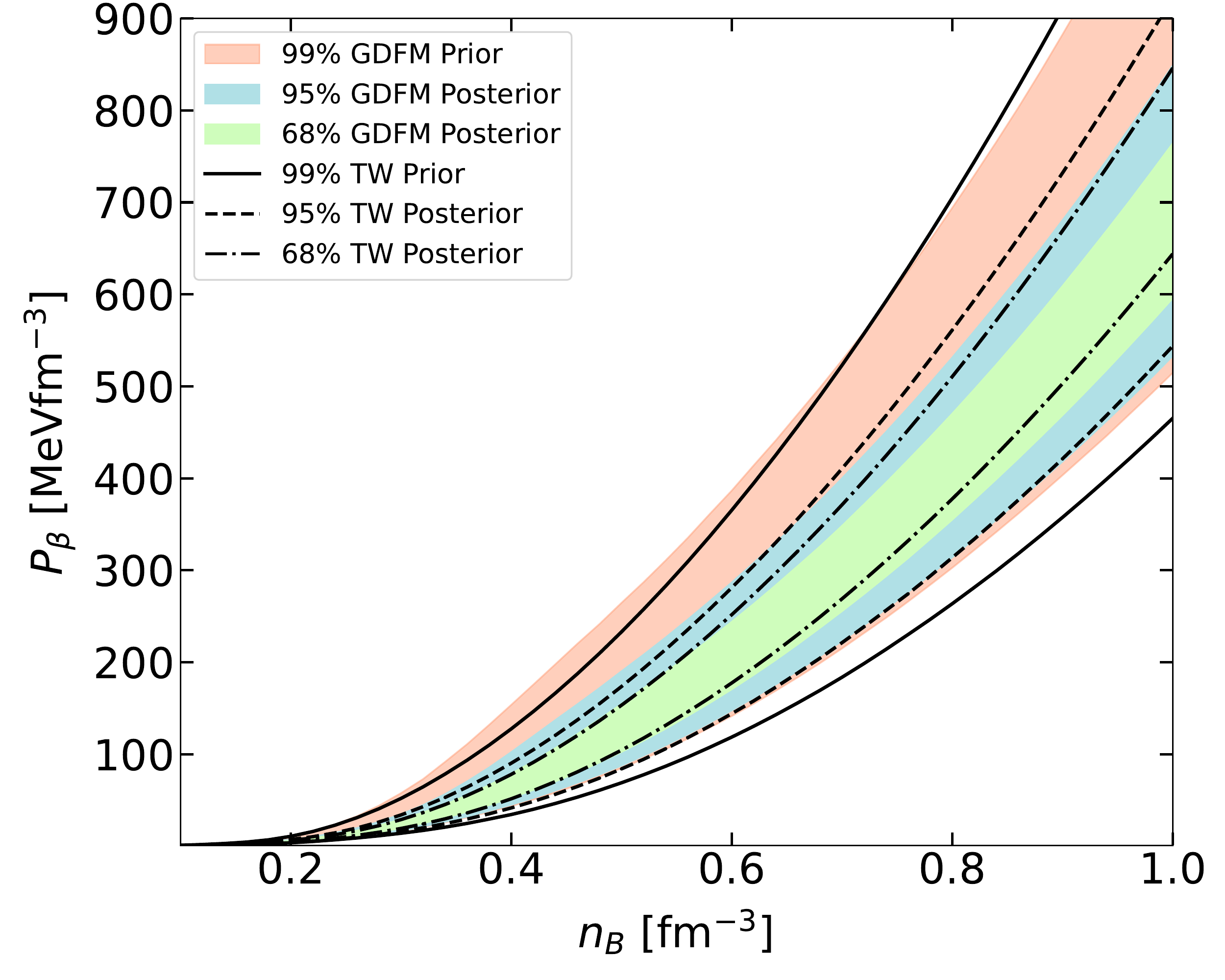}
\caption{{Contours of} pressure at $\beta$-equilibrium as a function of baryon density {at different CI for GDFM and TW metamodels}.}
    \label{fig:p-rho}
\end{figure}
{The correlations observed in the two-dimensional distributions of Figs. \ref{fig:isoscalar},\ref{fig:isovector} can be quantified by calculating the Pearson correlation coefficients. We have shown that for the priors (upper panel) and posteriors (lower panel) in Fig. \ref{fig:corr_NMP}.} 
GDFM values are shown on the lower left and TW values are shown on the upper right. We find that the coefficients are 
{different already at the level of priors. This is not surprising because our priors are informed by nuclear physics constraints. Furthermore, the difference in the prior correlations, particularly in the isovector sector, appear due to the different forms of Lagrangians for GDFM and TW.} Some prior correlations are reduced in the posterior. We find notable differences in the GDFM and TW models for the posterior Pearson coefficients between $K_{sat}$ and $Z^*_{sat}$, $Q^*_{sat}$; or $Z^*_{sat}$, $K_{sym}$ and $L_{sym}$, $L_{sym}$, $Z^*_{sym}$;  or $Q^*_{sym}$ and $Z^*_{sym}$. {These confirm the the different patterns shown in the two-dimensional distributions among different NMPs in Figs. \ref{fig:isoscalar} and \ref{fig:isovector}.}

In figure \ref{fig:p-rho}, we show the pressure of $\beta$-equilibrated matter as a function of baryon density for GDFM and TW for $99\%$ of the priors, and $68\%$ and $95\%$ of the posteriors, respectively. {Throughout the whole density range,} we find that the TW prior produces lower pressure consistently in comparison to GDFM. We also find that at lower densities, GDFM posterior is marginally stiffer than the TW, but the trend changes at higher densities when the TW becomes significantly stiffer. For both cases, astrophysical constraints rule out too stiff EOSs and the posteriors shrink 
{discarding} the higher pressure regions of the priors. TW posteriors at higher densities appear to {explore larger ranges of pressure} than those of GDFM.

\begin{figure}
    \centering
    \includegraphics[width=0.5\textwidth]{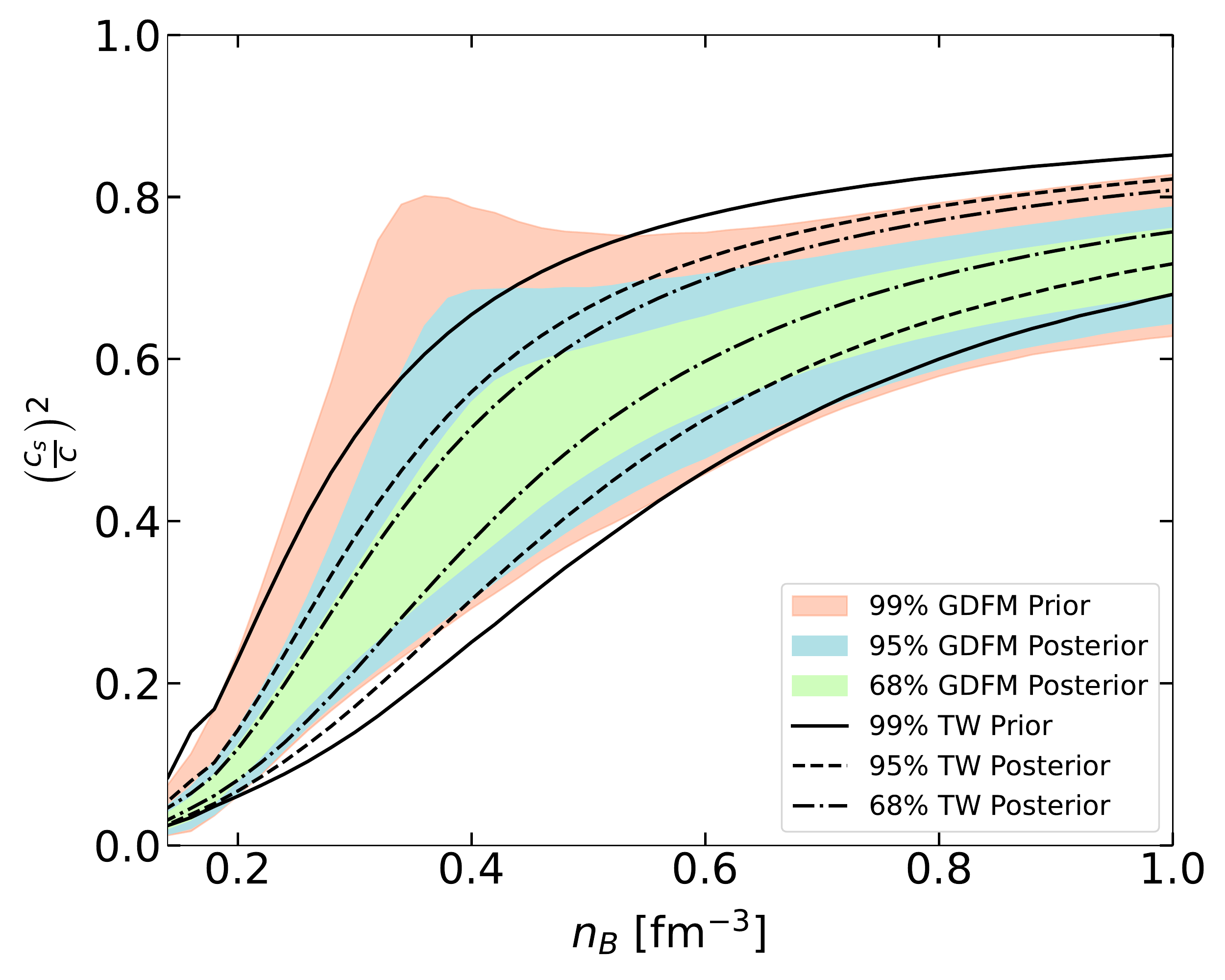}
\caption{Same as Fig. \ref{fig:p-rho}, but for speed of sound.}
    \label{fig:cs}
\end{figure}
The effect of the relative difference of stiffness at different densities can be clearly found in the speed-of-sound posteriors for GDFM and TW in Fig. \ref{fig:cs}. In Fig. \ref{fig:p-rho}, we have noticed that GDFM is relativly stiffer at low densities and softer at high densities. Consequently, we find in Fig. \ref{fig:cs} that the speed of sound for GDFM increases rapidly and 
{get saturated} with the increase of density. There {are} individual EOSs where the speed of sound actually decreases at high density. In case of TW, the speed of sound increases smoothly and stabilizes at higher densities. This apparent difference in bahavior can also be attributed to the isovector freedom in the model. The increase and decrease of the symmetry energy control the stiffness of the EOSs over the whole range of densities. This assertion becomes even clearer from the next figure where we plot the SNM and PNM properties from both GDFM and TW model. 

\begin{figure}
    \centering
    \begin{tabular}{c}
     \includegraphics[width=0.5\textwidth]
    {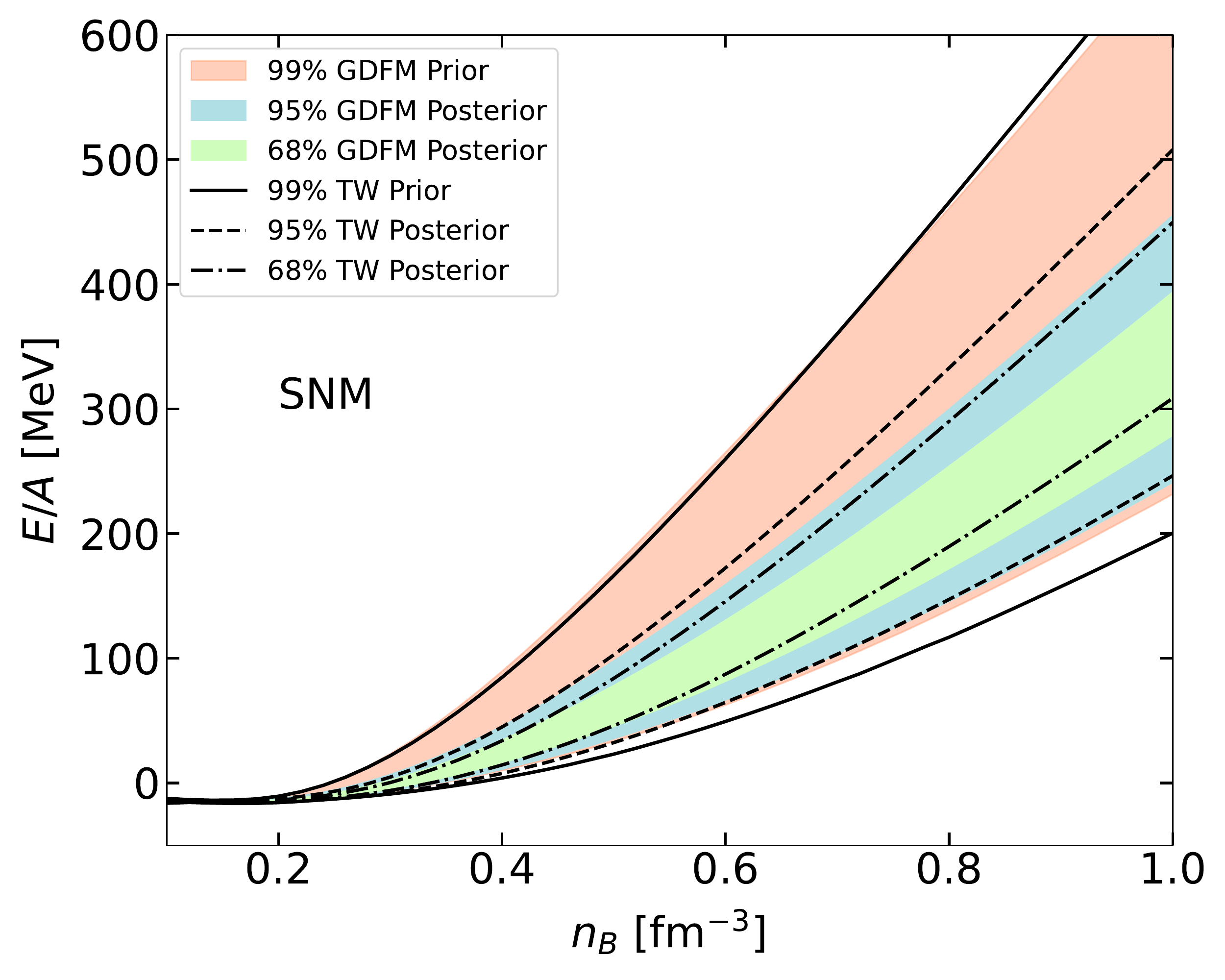} \\
    \includegraphics[width=0.5\textwidth]
    {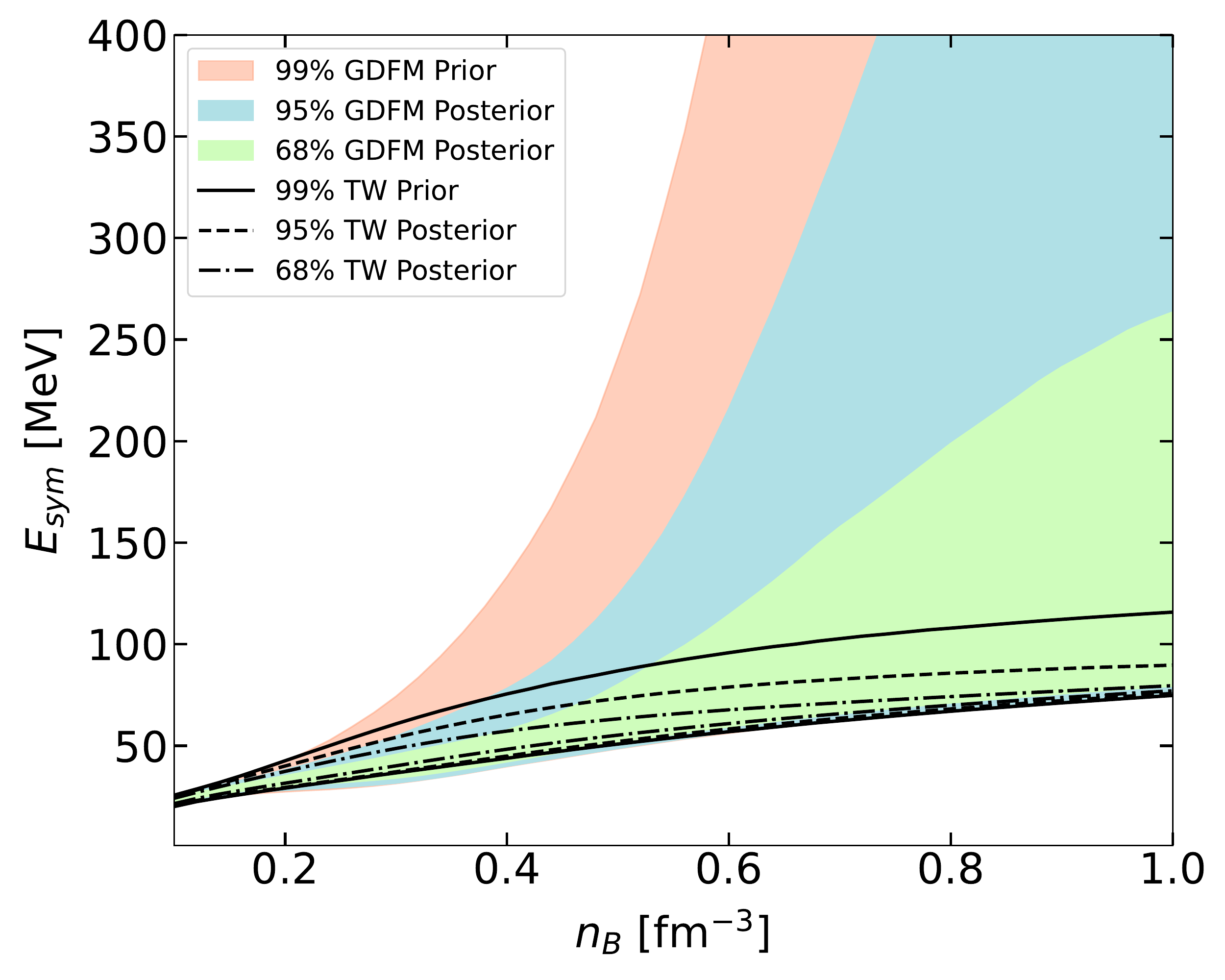}
    \end{tabular}
    \caption{Same as Fig. \ref{fig:p-rho}, but for energy per particle of the SNM (upper panel) and symmetry energy (lower panel).}
    \label{fig:SNM_PNM}
\end{figure}
{In Fig. \ref{fig:SNM_PNM}, we have shown the energy per particle for SNM (upper panel) and symmetry energy (lower panel) as a function of density. The lower boundary of the prior for SNM is slightly lower in TW. In the posteriors, the energy per baryon for SNM stays systematically higher for TW. They are also slightly wider. However, the behavior of the symmetry energy for TW and GDFM is completely opposite. The symmetry energy of GDFM is much wider than TW at all densities beyond 0.25 fm$^{-3}$, though, the lower bounds for TW and GDFM are almost identical. Again, we attribute this freedom to the isovector sector of the GDFM lagrangian, which also gives consistently larger ranges for the isovector NMPs (see Fig: \ref{fig:isovector}).}

\begin{figure}
    \centering
    \includegraphics[width=0.5\textwidth]{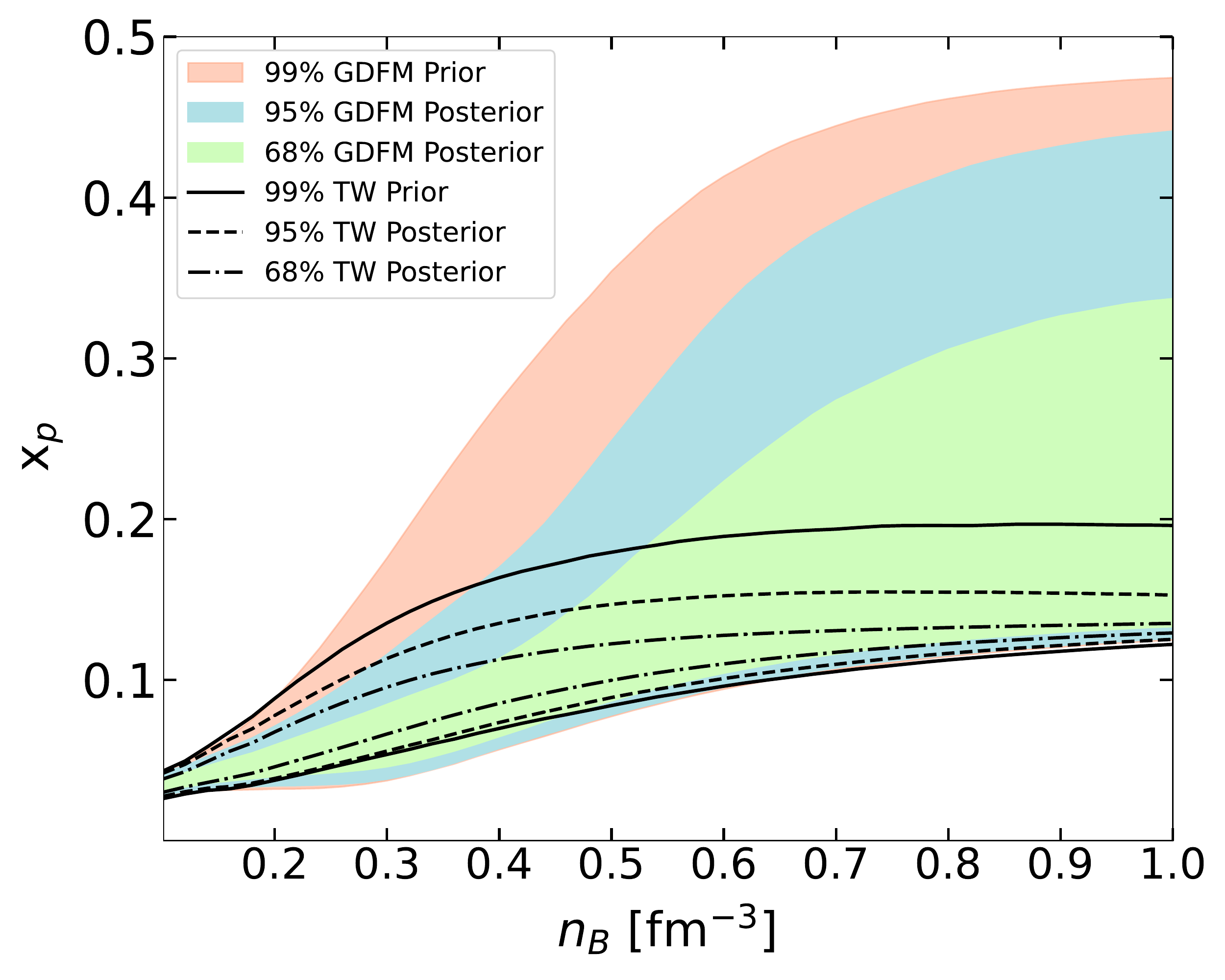}
\caption{Same as Fig. \ref{fig:p-rho}, but for proton fraction.}
    \label{fig:xp}
\end{figure}

\begin{figure}
    \centering
    \includegraphics[width=0.5\textwidth]{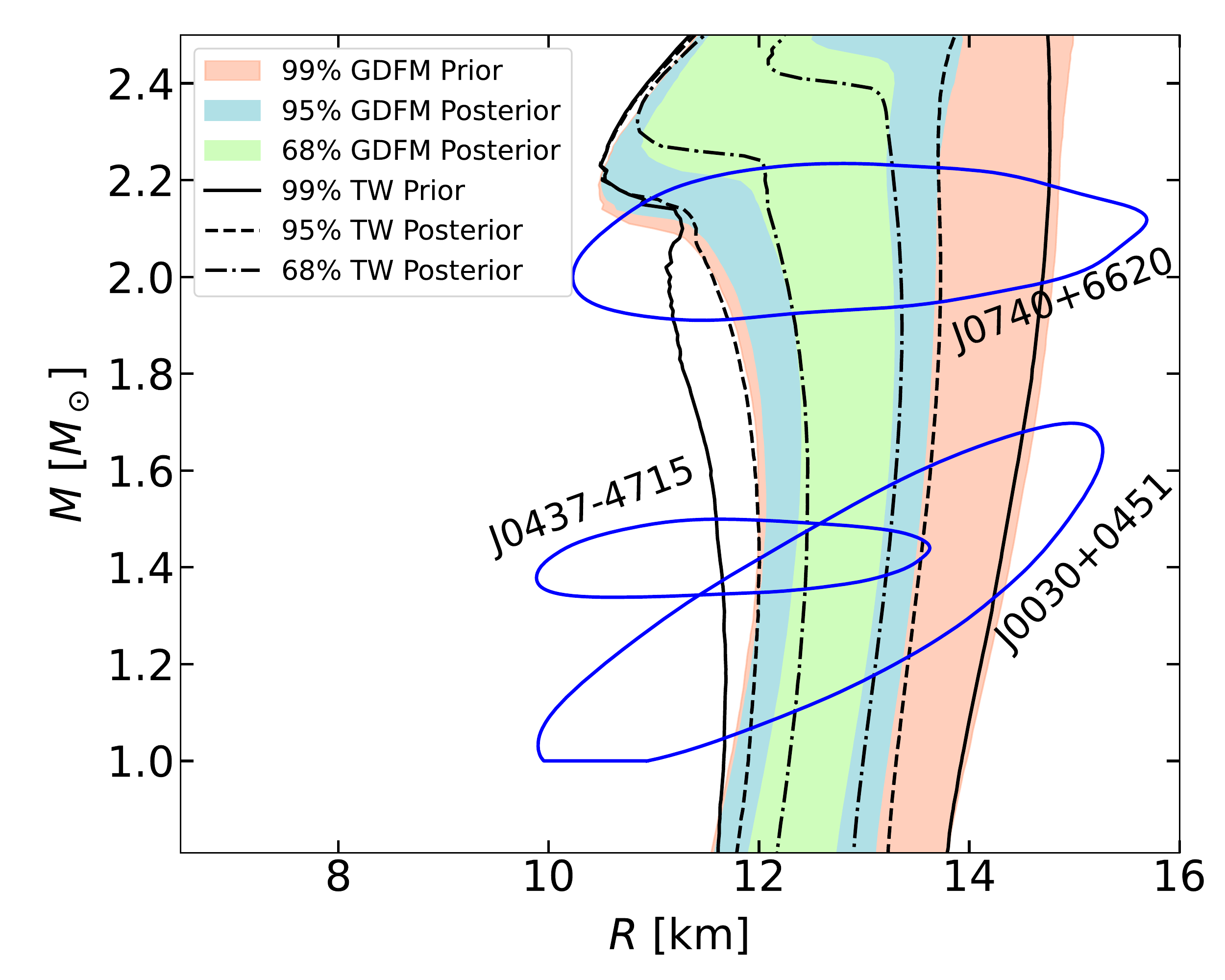}
\caption{Contours of mass-radius relations at different CIs corresponding to the EOS models shown in Fig.~\ref{fig:p-rho}. The solid blue contours represents the 95\% CI of the different NICER sources \cite{Riley:2019yda,Riley:2021pdl,Choudhury:2024xbk} (see text for details).}
    \label{fig:mr}
\end{figure}
\begin{figure}
    \centering
    \includegraphics[width=0.5\textwidth]{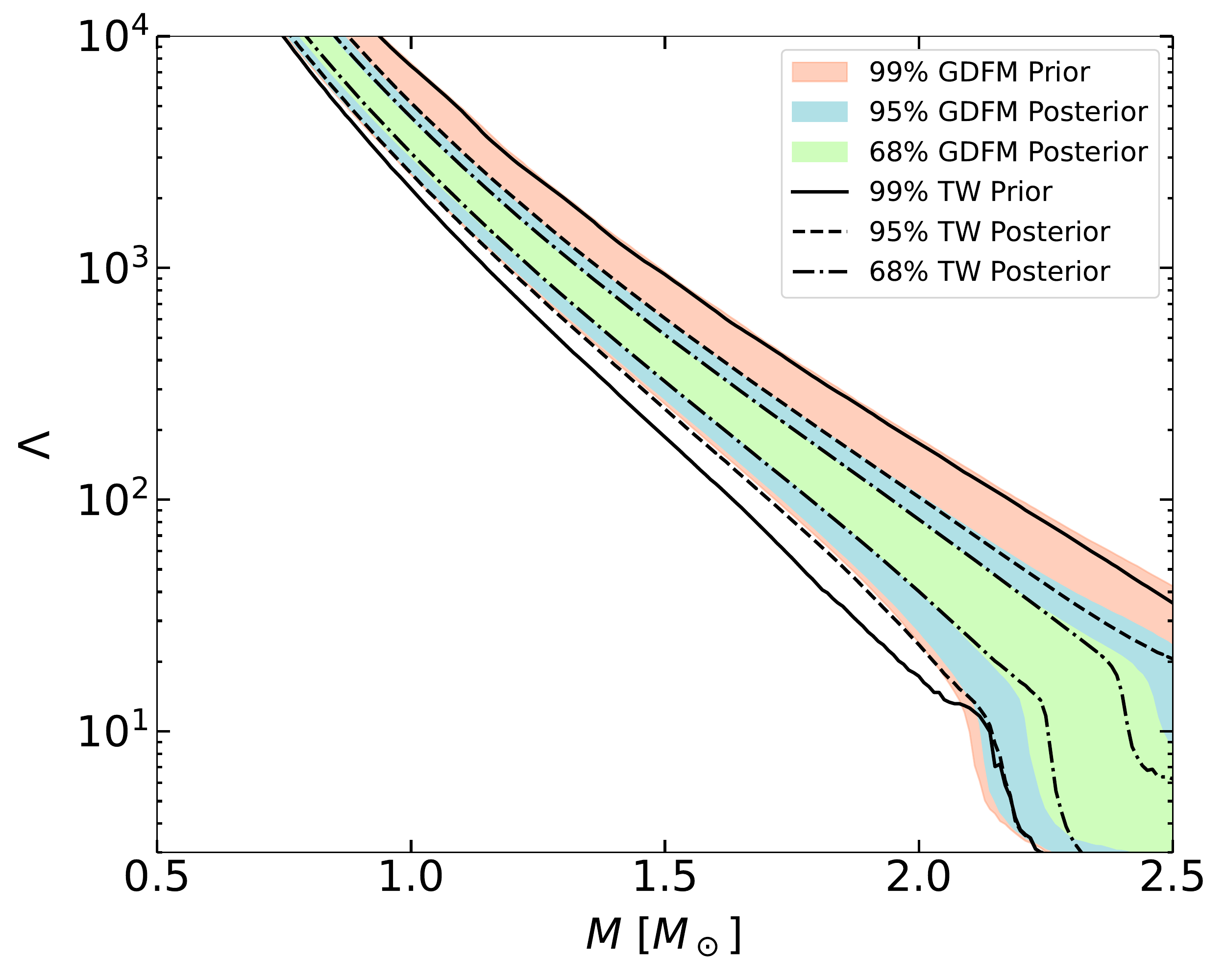}
\caption{Mass-tidal deformability relations corresponding to the EOS model ranges shown in Fig.~\ref{fig:p-rho}}
    \label{fig:mlambda}
\end{figure}
{The most significant feature of the freedom in the isovector sector mentioned above is manifested in the proton fraction contours in Fig. \ref{fig:xp}. The similar nature in SNM energy per particle but a contrasting one in the behavior of the desnity dependence of symmetry energy is mirrored here. We see that the {lower end of the} proton fraction for TW is very similar to that of GDFM. However, as density increases the proton fraction quickly saturates to a value between $0.12$ and $0.2$ for the prior. The posteriors become very narrow for TW at high densities, barely leaving any room to navigate if necessary for future experimental data. Next, we look at the GDFM. {Close to the crust region,} at lower densities, both the range and the width of the priors and posteriors are similar for TW and GDFM. But at higher densities, the width of the proton fraction increases rapidly for GDFM. It reaches as much as $\sim 0.45$ for the prior, also more than $\sim 0.3$ in the posterior. This behavior is fully consistent with the behavior of the EOS at $\beta$-equilibrium in Fig. \ref{fig:p-rho} and the speed of sound in Fig. \ref{fig:cs}; or the SNM and symmetry energy behavior as shown in Fig. \ref{fig:SNM_PNM}. 
 
It was shown in Ref. \cite{Mondal:2021vzt} that to pin the composition down at high densities, explicit information on the symmetry energy is essential. Looking at the behavior of symmetry energy in Fig. \ref{fig:SNM_PNM}, it is clear that due to larger freedom in GDFM, explored range in proton fraction by the same is significantly larger compared to TW. In this light, we anticipate a significantly different conclusion drawn in Ref. \cite{Carvalho:2023ele}, where it was conjectured that the composition can be extracted from $\beta$-equilibrated EOS using a machine learning approach. Incidentally, for training the neural network, a non-linear RMF model \cite{Furnstahl:1996wv, Todd-Rutel:2005yzo} was used in this work, which also explores very narrow range of proton fractions at high densities.}

\begin{figure*}
    \centering
    \includegraphics[width=0.8\textwidth]{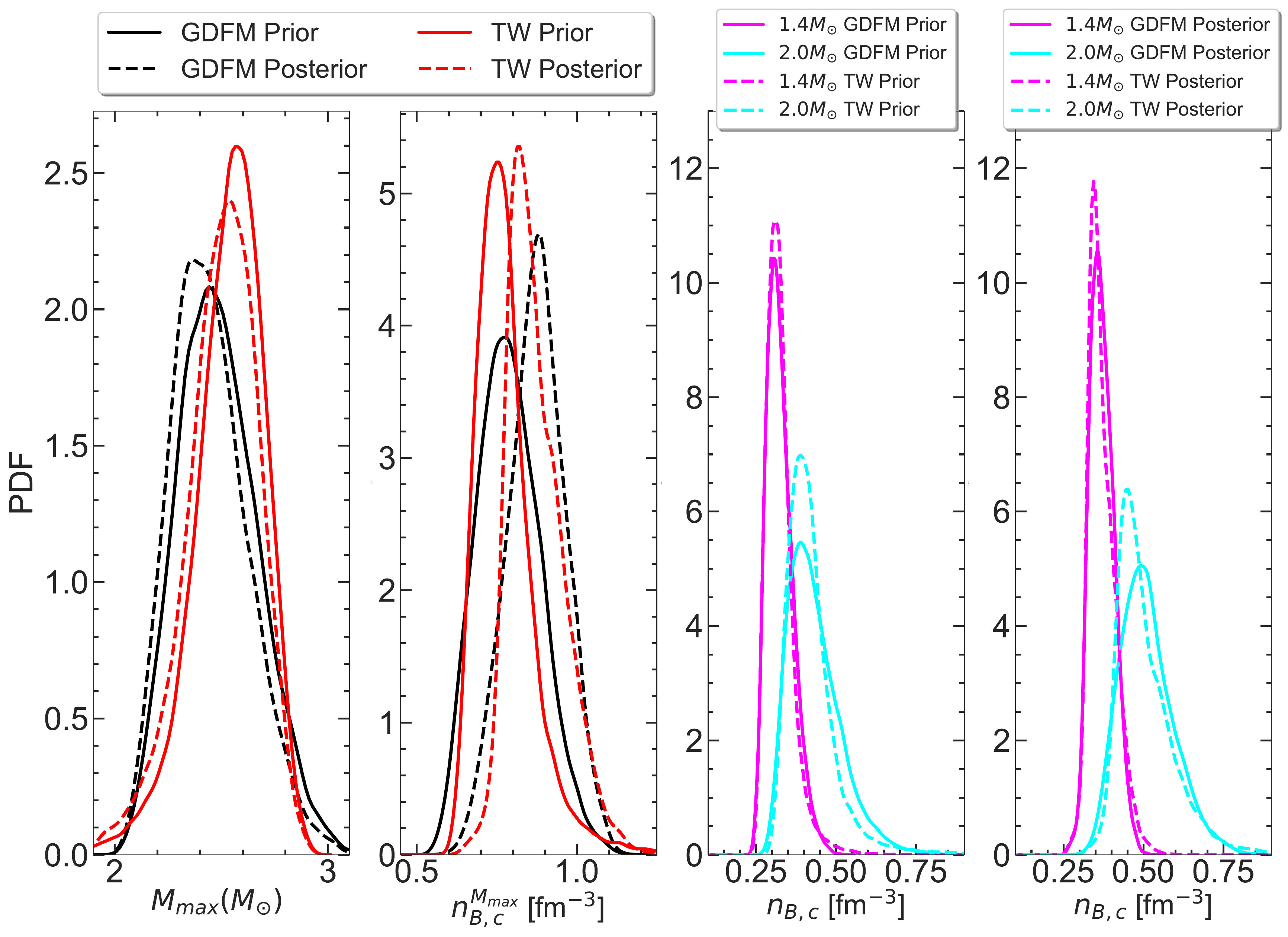}
    \caption{Distribution of maximum masses, central densities of maximum mass stars, the distributions of the central densities of $1.4M_\odot$ and $2.0M_\odot$ stars are shown for the GDFM and TW priors and posteriors, respectively.}
    \label{fig:mmax-central}
\end{figure*}
\begin{figure*}
    \centering
    \begin{tabular}{cc}
    \includegraphics[width=0.5\textwidth]{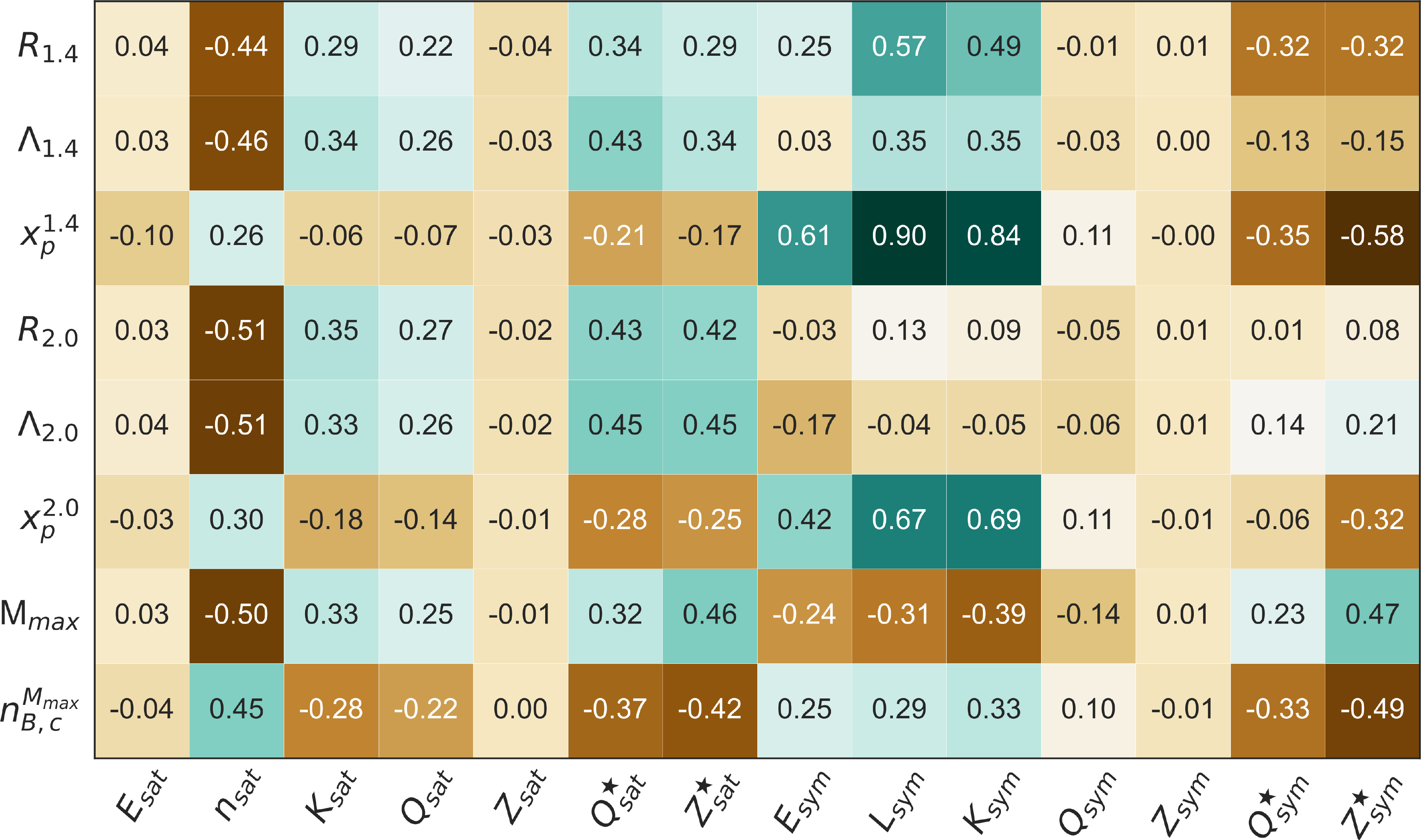}
    \includegraphics[width=0.5\textwidth]{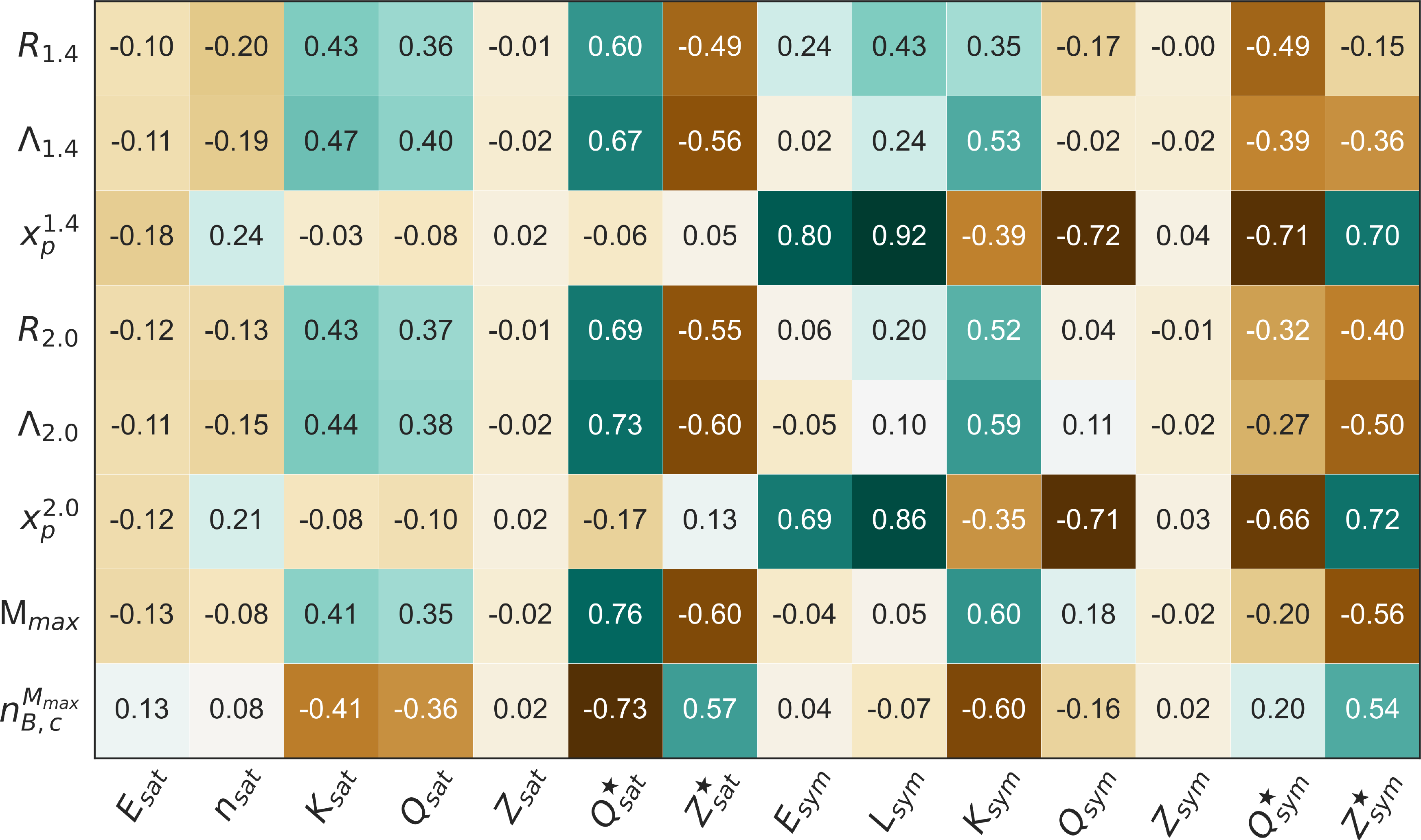} 
    \end{tabular}
    \caption{{Pearson correlation coefficients among NMPs and some selected NS properties for GDFM (left panel) and TW (right panel) posteriors.}}
    \label{fig:corr_NMP_NS_post}
\end{figure*}
Next, we focus on the neutron star structure emerging form GDFM and TW functionals. In Figs. \ref{fig:mr} and \ref{fig:mlambda}, we have shown the $M-R$ and $M-\Lambda$ sequences, respectively. In Fig. \ref{fig:mr}, we have also shown the $95\%$ contours for three NICER sources \cite{Riley:2019yda,Riley:2021pdl,Choudhury:2024xbk}. We can see that the posteriors for both GDFM and TW are consistent with the NICER observations. As before, the prior contours for both models are from the EOSs informed purely by nuclear physics knowledge. We find that the TW model produces slightly wider prior {exploring more} in the lower radii than the GDFM. Both of them consistently produce large radii which are then excluded in the posteriors due to the tidal deformability constraint. Similarly, {lower} tidal deformability values  
{are probed in TW than in GDFM, as can be seen} in Fig. \ref{fig:mlambda}. This behavior is consistent with the finding of previous literatures \cite{Malik:2022zol, Char:2023fue, Scurto:2024ekq}. The posteriors {for TW and GDFM} look very similar 
both at $68\%$ and $95\%$ CI. This is again consistent from what is expected from the EOS posteriors in Fig. \ref{fig:p-rho}. However, a closer look at Fig. \ref{fig:mr}  
{can} reveal some subtle differences between TW and GDFM also consistent with their features from Fig. \ref{fig:p-rho}. For example, the TW prior produces lower radii for low-mass star, but slightly larger radii for high mass stars compared to GDFM. For the posterior of TW, we find that it is narrower for lower masses and very high masses, but around $1.8$ - $2.0$ M$_\odot$, it almost overlaps with the one from GDFM. This is corresponding to the central densities $0.4$ - $0.6$ fm$^{-3}$ for which, we see an overlap of the TW and GDFM posteriors, also in the EOS in Fig. \ref{fig:p-rho}. 

{In Fig. \ref{fig:mmax-central}, we have shown the distribution of maximum mass of neutron stars, central densities of maximum mass ($n_{B,c}^{M_{max}}$) and the central densities for $1.4M_\odot$ and $2.0M_\odot$ for the prior and posterior distributions of GDFM and TW models. Both for GDFM and TW, the $n_{B,c}^{M_{max}}$ peaks shift to higher values from prior to posterior. Interestingly, the priors are very similar. The same later feature is also seen for $n_{B,c}^{M_{1.4\odot}}$ and $n_{B,c}^{M_{2.0\odot}}$ (right two panels of Fig. \ref{fig:mmax-central}). Overall, the densities explored in GDFM and TW are not very different, which is in accordance with the behavior of $\beta$-equilibrated matter (see Figs. \ref{fig:p-rho}, \ref{fig:mr}, \ref{fig:mlambda}), even though composition  (\ref{fig:xp}) or symmetry energy as a function of density (\ref{fig:SNM_PNM}) can be quite different. }

{To understand further the differences explored by GDFM and TW posteriors, in Fig. \ref{fig:corr_NMP_NS_post} we plot the Pearson correlation coefficient between the NMPs and a few chosen neutron star properties for GDFM in the left panel and for TW in the right panel. We focused particularly on the mass and the tidal deformability of $1.4M_\odot$ and $2.0M_\odot$ stars, their proton fraction at the central density ($x_p^{1.4}$ and $x_p^{2.0}$, respectively) and the maximum mass and central density corresponding to the maximum mass. Major differences in the correlations among neutron star properties and $n_{sat}, Q^*_{sat}, Z^*_{sat}, K_{sym}, Q^*_{sym}, Z^*_{sym}$ can be observed between the two panels. This can clearly be attributed to the difference in the Lagrangian for GDFM and TW, which also induces the major difference in the behavior of symmetry energy and proton fraction as a function of density, as seen in Figs. \ref{fig:SNM_PNM}, \ref{fig:xp}. For $x_p^{1.4}$ and $x_p^{2.0}$, one can observe a curious correlation with $Q_{sym}$ in the TW case, even though the latter is sampled independently to calculate the crust. Upon minute observation from the lower panel of Fig. \ref{fig:corr_NMP}, one can conclude that this correlation comes through the correlation of $Q_{sym}$ with $L_{sym}, Q^*_{sym}$ and $Z^*_{sym}$, which control the composition through symmetry energy at high densities. These correlations are totally absent in GDFM.}

\section{Conclusion}
In the present work, we have investigated two density-dependent RMF models, TW and GDFM \cite{Typel:1999yq,Gogelein:2007qa}, within a metamodelling approach developed in Ref. \cite{Char:2023fue}. We were able to generate a wide range of NMPs from both models with our choice of the coupling parameters. As the first step, we have performed a Bayesian inference for the model parameters with Nested Sampling using the constraints on the NMPs from Table \ref{tab:prior} and $\chi$-EFT \cite{Huth:2020ozf} to get an optimized space for coupling parameters and subsequently the NMPs derived from them. This is an improvement over our previous work, where random sampling was used to create a large sample of model parameters satisfying those constraints. Then, we have calculated the low-density EOS with the same set of NMPs for both TW and GDFM, following \textcite{Carreau:2019zdy}.  We have developed a unified EOS prescription with the low and the high density parts of the EOSs are joined at the saturation. 

{This procedure was adapted to get rid of numerical inconsistencies at crust-core transition density}. The benefit of using relativistic density functionals {at higher densities} is that the EOS remains causal throughout the range. With those unified EOSs at hand, we have applied the astrophysical constraints on NS mass, radius, tidal deformabality, in a Bayesian way. As we have seen before \cite{Char:2023fue}, the constraints from GW170817 and $\chi$-EFT are more effective in modifying the priors for both GDFM and TW, while the NICER results have a minimal effect. We have found that lower order NMPs have a similar range for both GDFM and TW, however, their correlations differ significantly. The difference is starker for the higher order NMPs. This leads to our most significant findings on the proton fraction produced by the two models. Although GDFM produces a large variation in the proton fraction, it is very narrow for TW. We found the reason behind this behavior to be associated with the freedom in the isovector part of the effective Lagrangian of GDFM functional, and consequently the freedom in the density dependence of symmetry energy. This deviation in the composition of the NS matter can have significant implications on the NS cooling and other transport behaviors. In addition, we have found the posterior contours $M-R$ and $M-\Lambda$ to be very similar for both the GDFM and the TW models. This substantiates previous conclusions found in the literature \cite{Mondal:2021vzt,Xie:2020tdo} that the composition of NS matter cannot be probed by observation of NS properties that are the result of matter at $\beta$-equilibrium.

\section*{Acknowledgements}
The authors thank Micaela Oertel, Francesca Gulminelli, and Tuhin Malik for helpful discussions and comments. This project has received funding from the European Union’s Horizon 2020 research and innovation programme under the Marie Skłodowska-Curie grant agreement No. 101034371. PC acknowledges the support from the European Union's HORIZON MSCA-2022-PF-01-01 Programme under Grant Agreement No. 101109652, project ProMatEx-NS. CM acknowledges partial support from the Fonds de la Recherche Scientifique (FNRS, Belgium) and the Research Foundation Flanders (FWO, Belgium) under the EOS Project nr O022818F and O000422.

\bibliography{mybiblio}
\end{document}